\documentclass{amsart}

\usepackage{a4wide}
\usepackage{xspace}
\usepackage{graphicx}
\usepackage{segbib2} 
\usepackage{amsmath,amsfonts}

\newcommand{\comment}[1]{}

\providecommand{\ensuremath}[1]{\ifmmode#1\else$#1$\,\fi}

\newcommand{\sigle}[1]{\textsc{#1}\xspace}

\newcommand{\MKSA}{\sigle{MKSA}}

\newcommand{\GPR}{\sigle{GPR}}

\newcommand{\sig}{\ensuremath{\sigma}\xspace}

\newcommand{\eps}{\ensuremath{\varepsilon}\xspace}

\renewcommand{\i}{\ensuremath{\rm i}}

\newcommand{\vecteur}[1]{{\ensuremath{\rm \bf{#1}}}\xspace}

\newcommand{\nul}{\vecteur{0}}

\newcommand{\E}{\vecteur{E}}
\newcommand{\B}{\vecteur{B}}

\newcommand{\J}{\vecteur{J}}

\newcommand{\x}{\vecteur{x}}

\renewcommand{\r}{\vecteur{r}}
\newcommand{\R}{\vecteur{R}}

\let\dt\partialt

\let\dttwo\partialttwo
\newcommand{\rot}{\vecteur{\nabla} \times}
\newcommand{\dive}{\vecteur{\nabla} \cdot}

\newcommand{\unite}[1]{\text{#1}}

\newcommand{\metre}{\unite{m}}

\newcommand{\henri}{\unite{H}}
\newcommand{\fahrad}{\unite{F}}

\let\m\metre

\let\S\siemens

\title{GROUND PENETRATING RADAR:\\
ANALYSIS OF POINT DIFFRACTORS FOR MODELING AND INVERSION}

\author{Albane Saintenoy}
\address{Institut de Physique du Globe, 
         4, place Jussieu, 75252 Paris Cedex 05, France\\
         Colorado School of Mines,
         Department of Geophysics, Golden CO 80403, USA}

\author{Albert Tarantola}
\address{Institut de Physique du Globe, 
         4, place Jussieu, 75252 Paris Cedex 05, France}

\begin{document}
\maketitle


\begin{abstract}

The three electromagnetic properties appearing in Maxwell's equations
are dielectric permittivity, electrical conductivity and magnetic
permeability. The study of point diffractors in a homogeneous,
isotropic, linear medium suggests the use of logarithms to describe
the variations of electromagnetic properties in the earth. A small
anomaly in electrical properties (permittivity and conductivity)
responds to an incident electromagnetic field as an electric dipole,
whereas a small anomaly in the magnetic property responds as a
magnetic dipole. Neither property variation can be neglected without
justification. Considering radiation patterns of the
different diffracting points, diagnostic interpretation of electric
and magnetic variations is theoretically feasible but is not an easy
task using Ground Penetrating Radar. However, using an effective
electromagnetic impedance and an effective electromagnetic velocity to
describe a medium, the radiation patterns of a small anomaly behave
completely differently with source-receiver offset. Zero-offset
reflection data give a direct image of impedance variations while
large-offset reflection data contain information on velocity
variations.

\end{abstract}


\section{Introduction}
\label{sec:intro}

Ground Penetrating Radar (\GPR) data yield information on the electric
and magnetic properties of a medium with good resolution (from a few
centimeters for a 900~MHz antenna to a few meters for a 50~MHz
antenna). A key issue is finding a good parameterization of the
subsurface for the inverse problem, including earth media containing
high magnetic permeability perturbations (ferrous metallic objects, magnetite,
iron-bearing rocks,...).

It is now well understood \cite{tarant86,debski95} that, when using
multi-offset seismic data, from an elastic medium one can resolve, in
order of importance, contrasts in acoustic impedance (from the
reflection amplitude at small offsets), contrasts in Poisson's ratio
(from the variation of reflection amplitude as a function of offset),
and contrasts in mass density (being poorly resolved). These elastic
parameters are nonlinear combinations of the Lam\'e parameters that
appear explicitly in the elastic wave equation. When we started the
present research, it was not obvious which functions of the three
electromagnetic parameters (dielectric permittivity, electrical
conductivity, and magnetic permeability) could be resolved using \GPR
data. Following a seismic approach \cite{tarant86}, we show that
the effective electromagnetic impedance and the effective
electromagnetic velocity can be resolved.

Modeling of \GPR data requires solving Maxwell's equations. One
approach is to linearize Maxwell's equations by approximating the
medium as a superposition of point diffractors superimposed on a
smooth surrounding medium. Here we study the behavior of a single
point diffractor in an isotropic, linear, homogeneous medium, excited
by a propagating electromagnetic wave. The response of such a point
diffractor can be found using the Stokes and Mueller matrix
considering only the electrical properties \cite{ulaby90}. We do not
limit our study to the electrical properties only, we also include the
magnetic properties and a magnetic point diffractor.

In addition to being relevant for inversion and modeling, a
point diffractor analysis can be useful when trying to characterize an
antenna radiation pattern \cite{arcone95,rossiter88}. The experiment
consists of recording monostatic radar data over a small object
in a lake or in the ground. Wrongly assuming an isotropic
response from small objects adds to the uncertainty in the radiation
pattern interpretation of those data. In the following we derive a
first order analytical expression for the field diffracted by point
diffractors, assuming small logarithmic parameter perturbations.

\section{Point diffractors}

\subsection{Logarithmic parameters}

\begin{figure}
\centerline{\includegraphics*[width=12cm]{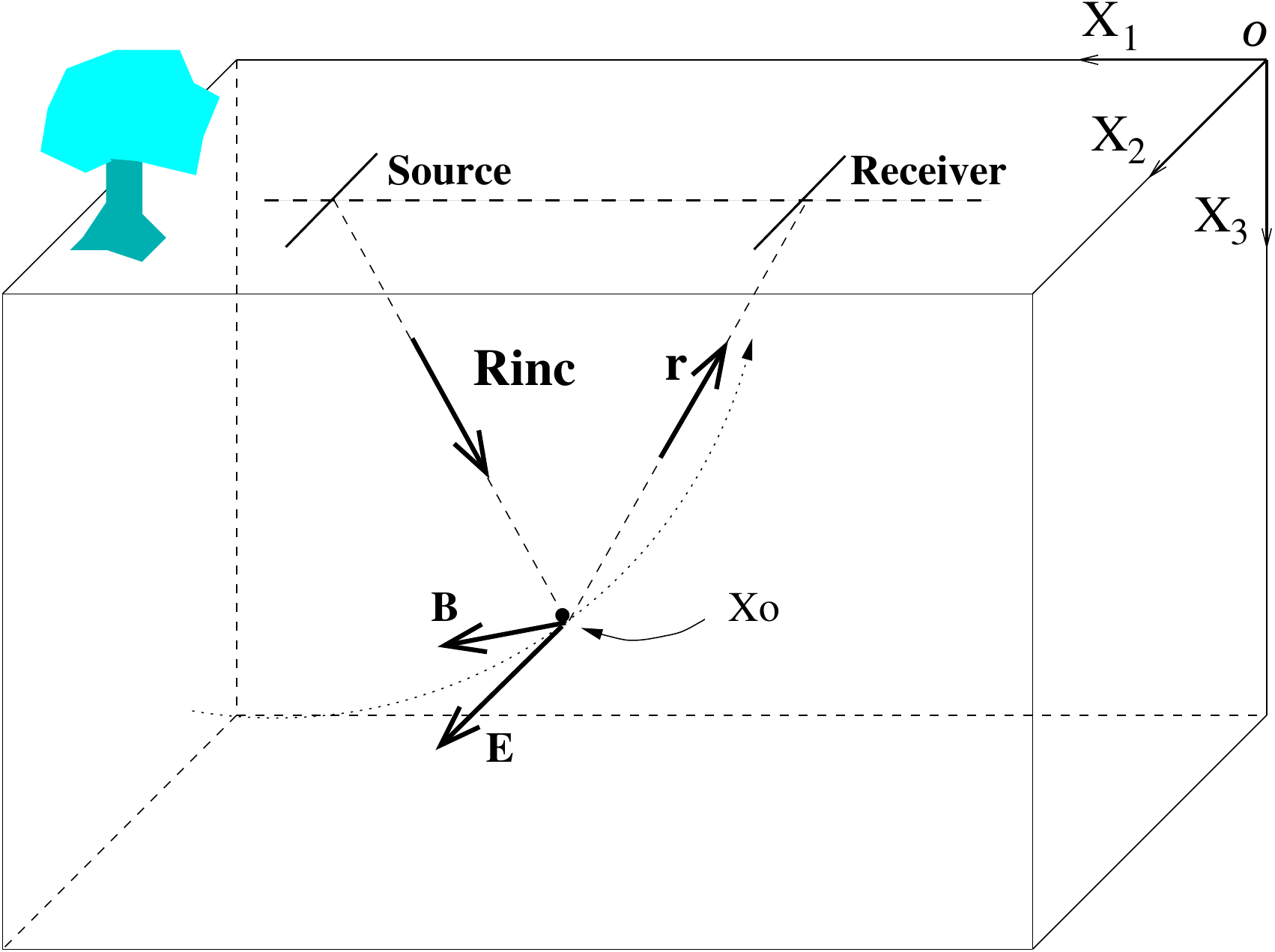}}
\vspace{5mm}
\caption{Position of the antennas and polarization of incident
electromagnetic fields \E and \B when arriving at the anomaly at point
$\protect\x_0$.}
\label{fig: champ incident}
\end{figure}

\GPR sends an electromagnetic wave into the ground and measures the
amplitude of the back-scattered electric field. This wave consists of
a coupled pair of electric field \E and magnetic field \B. An external source,
$\J_s$ (e.g. the current density created by a bow-tie antenna), is
present in three-dimensional space $R^3$. To specify position, we use
the coordinates {$x_1$, $x_2$, $x_3$} with respect to an orthogonal,
Cartesian reference frame with origin $O$ and three mutually
perpendicular base vectors of unit length forming a right-handed
system (Figure~\ref{fig: champ incident}). When appropriate, the
space coordinates are collectively denoted by the position vector \x.
The time coordinate is denoted by $t$.

In response to $\J_s$, currents result from polarization caused by
displacement and conduction (Ohm's law) inside the explored medium.
Electromagnetic fields in an isotropic (so parameters are scalar),
linear medium with time-independent parameters, are governed by
Maxwell's equations \cite{jackson98}, which in the \MKSA system are,
\begin{equation}
\dive \left(\eps({\x}) \E ({\x},t)\right)\ =\ \rho ({\x},t),
\end{equation}
\begin{equation}
\rot \left(\frac{\B({\x},t)}{\mu({\x})}\right)\
- \eps({\x}) \dt \E ({\x},t)\ -\ \sig({\x})
\E ({\x},t)\ =\ {\J}_s({\x},t),
\label{eq: rot H}
\end{equation}
\begin{equation}
\dive \left(\frac{\B({\x},t)}{\mu({\x})}\right)\ =\ 0,
\end{equation}
\begin{equation}
\rot \E({\x},t)\ +\ \dt \B({\x},t)\ =\ \nul,
\label{eq: rot E}
\end{equation}
where $\rho$ is the density of electric charges, $\nabla$ is the
partial differential operator [$\nabla^T=(\frac{\partial}{\partial
  x_1}, \frac{\partial}{\partial x_2}, \frac{\partial}{\partial
  x_3})$], and $\times$ indicates the cross product. The parameters
appearing in these equations are dielectric permittivity \eps,
electrical conductivity \sig, and magnetic permeability $\mu$, which
are all positive. For earth materials
\cite{olhoeft79,olhoeft93,schon96},
\begin{equation}
\left\{\begin{array}{ccl}
{\eps}_0              & \leq \eps   & \leq 100\ {\eps}_0,\\
10^{-7}\,\S/\m               & \leq  \sig   & \leq 10^7\,\S/\m,\\
(1\,-\,10^{-4})\,\mu_0   & \leq \mu    & \leq 100\,\mu_0.
\end{array}\right.
\label{eq: inegalites}
\end{equation}
The reason that the dielectric permittivity \eps is always greater than
$\eps_0$ (the dielectric permittivity of vacuum), is explained by
Landau and Lifshitz \shortcite{landeau60} from thermodynamic
considerations. Rocks presenting exclusively diamagnetic properties
are characterized by a magnetic permeability lower than $\mu_0$ (the
magnetic permeability in vacuum). For paramagnetic, ferro and
ferri-magnetic rocks, $\mu > \mu_0$.

Parameters in inequalities~(\ref{eq: inegalites}) are positive and
non-zero parameters. This allows the definition of logarithmic
parameters $\eps^*$, $\sig^*$ and $\mu^*$ (logarithm of the linear
parameter over an arbitrarily chosen reference value) as follows,
\begin{equation}
{\eps}^* = \ln\left(\frac{\eps}{{\eps}_0}\right),
\end{equation}
\begin{equation}
{\sig}^* = \ln\left(\frac{\sig}{{\sig}_0}\right),
\end{equation}
and
\begin{equation}
\mu^* = \ln\left(\frac{\mu}{\mu_0}\right),
\end{equation}
where $\mu_0 = 4 \pi\ 10^{-7}\,\henri/\metre$, $\eps_0 \approx 8.854\ 
10^{-12}\, \fahrad/\metre$, and $\sig_0$ is chosen to be 1 S/m. In the
following, point diffractors will be contrasts in these
time-independent logarithmic parameters and we will assess their
implication upon the acquisition geometry of \GPR data.

\subsection{Diffracted field using (\eps, \sig, $\mu$) parameterization}

Equation~(\ref{eq: rot H}) defines, in part, the behavior of the
incident electromagnetic field created by $\J_s$ propagated in a
homogeneous, linear and isotropic medium characterized by
time-independent electromagnetic parameters \eps, $\mu$ and \sig. With
the presence of a small anomaly, the perturbed medium parameters
become $\eps + \delta \eps$, $\sig + \delta \sig$, and $\mu + \delta
\mu$. The total field propagated in the perturbed medium is the sum of
the incident field and the field scattered by the anomaly, $\E +
\delta \E$ and $\B + \delta \B$.

We calculate the Fr\'echet, or functional, derivatives of the
wavefield $\delta \E$ with respect to the medium parameters. We
compute them by taking the first order of a series expansion. For
example, the derivative of $\sin(x)$ is computed by expanding $\sin(x\ 
+\ \delta x)\ =\ \sin(x)\ +\ \cos(x) \delta x\ +\ ...$ and keeping
only the first order term [i. e., the derivative of $\sin(x)$ is
$\cos(x)$]. 

An anomaly in linear parameters, $\delta m$, is related to an anomaly
in logarithmic parameters $\delta m^*$ by
\begin{equation}
\delta m^*\ =\ (m\ +\ \delta m)^*\ -\ m^*\ =\ \ln \frac{m\ +\ \delta
m}{m_0}\ -\ \ln \frac{m}{m_0}\ =\ \ln \frac{m\ +\ \delta m}{m},
\end{equation} 
where $m$ is either \eps, \sig, or $\mu$. It follows that
\begin{equation}
m\ +\ \delta m = m \exp \delta m^* \approx  m (1\ +\ \delta m^*).
\label{eq: var linearisees}
\end{equation}
The interpretation of this first order approximation is examined below.

After canceling the background terms and keeping only first order
terms, the perturbed parameters~(\ref{eq: var linearisees}) in equation
(\ref{eq: rot H}) yield 
\begin{equation}
\begin{split}
{\rot} \left(\frac{\delta \B({\x},t)}{\mu}\right)\
& - {\eps}\ \dt \delta \E ({\x},t)\ -\ {\sig}\ \delta \E({\x},t)\ =\\
& {\rot} \left(\frac{\delta {\mu}^*({\x})}{\mu}\ \B ({\x},t) \right)\ +\ 
{\eps}\ \delta {\eps}^*({\x})\ \dt \E({\x},t)\ +\ {\sig}\ \delta {\sig}^*({\x})\
\E({\x},t).
\label{eq: res1}
\end{split}
\end{equation}
It is as if the perturbed fields, $\delta \E$ and $\delta \B$,
propagate in the non perturbed medium described by parameters \eps,
\sig and $\mu$, and originate from virtual electric sources that are
the terms on the right side in equation~(\ref{eq: res1}). Those
secondary sources depend on logarithmic parameter variations and on
the electromagnetic fields \E and \B that would have existed if there
were no perturbations. The anomaly behaves like a secondary electric
source that scatters $\delta \E$ and $\delta \B$.

When the perturbation in parameters is localized over a volume, $V$,
centered at point ${\x}_0$, the logarithmic contrasts can be written
as  
\begin{equation}
\delta {\eps}^*({\x}) = A_{{\eps}}^*\ \delta_{1/V}({\x} - {\x}_0),
\label{eq: contrasts 1}
\end{equation}
\begin{equation}
\delta {\sig}^*({\x}) = A_{\sig}^*\ \delta_{1/V}({\x} - {\x}_0),
\label{eq: contrasts 2}
\end{equation}
and
\begin{equation}
\delta \mu^*({\x}) = A_{\mu}^*\ \delta_{1/V}({\x} - {\x}_0),
\label{eq: contrasts 3}
\end{equation}
where $\delta_{1/V}({\x} - {\x}_0)$ is a smooth function that
converges to the Dirac distribution $\delta (\x - \x_0)$ as $V
\rightarrow 0$. The term $A_{\eps}^*$ is a perturbation in the
logarithmic permittivity multiplied by the perturbation volume $V$. It
is the same with $A_{\sig}^*$ and $A_{\mu}^*$ for the logarithmic
conductivity perturbation and the logarithmic permeability
perturbation, respectively.

The question now becomes, what are the characteristics of the electric
field diffracted by such a point source? When
${\J}_s({\x},t)\,=\,{\J}(t) \delta({\x}-{\x}_0)$, and $\J(t)$ is a
density of currents independent of spatial position, there is creation
of an electric field that is given by solving
\begin{equation}
\nabla \nabla \cdot \E ({\x},t)\ -\ \nabla^2 \E({\x},t)\ -\ \mu {\eps}
\dttwo \E ({\x},t) \ +\ \mu {\sig} \dt \E ({\x},t)\ =\ -\mu \dt {\J}_s({\x},t).
\label{eq: eq. des ondes}
\end{equation}
This equation is the curl of equation~(\ref{eq: rot E}) combined with
the derivative of equation~(\ref{eq: rot H}) with respect to time. The
solution to equation~(\ref{eq: eq. des ondes}) is a Green's tensor
whose ($p,q$) component in the frequency domain is
\begin{equation}
G^{pq}({\x}, \omega, \x_0)\ =\ \frac{\exp(\i \omega r/c)}{4 \pi r} (\delta^{pq}\ -\
\gamma^p \gamma^q)\ +\ \frac{\i c}{\omega} \frac{\exp(\i \omega r/c)}{4
\pi r^2} (1\ +\  \frac{\i c}{\omega r}) (\delta^{pq}\ -\
3 \gamma^p \gamma^q),
\label{eq: green frequence}
\end{equation}
where $p\,\in\,\{1,2,3\}$ and is the resulting electric field
direction index, $q\,\in\,\{1,2,3\}$ and is the source direction
index, $\i\ =\ \sqrt{-1}$, $\omega$ is the frequency,
$r\,=\,||{\x}-{\x}_0||$ is the distance between the observation point
$\x$ and the source position $\x_0$, $c=\sqrt{1/\mu (\eps + \i
\sig/\omega)}$ is the electromagnetic wave speed, and
$\gamma^p\,=\,(x^p-x_0^p)\,/\,r$. Details on the development of this
expression can be found in \cite{saintenoy98}.

When the medium has low conductivity (${\sig} \ll {\eps} \omega$), and
the second term in equation~(\ref{eq: green frequence}) is small
compared to the first one ($r \gg \lambda$ where $\lambda$ is the
wavelength), the far-field term of the ($p,q$) component of the
Green's tensor is written in the space-time domain as
\begin{equation}
G^{pq}({\x}, t, {\x}_0, t_0)\ =\ \frac{1}{4 \pi r}(\delta^{pq}\ -\
\gamma^p \gamma^q) \delta(t\ -\ t_0\ -\ \frac{r}{c}).
\label{eq: green non conducteur}
\end{equation}
The $p$-th component of the far-field term of the electric field
diffracted when ${\J}_s({\x},t)$ is more general and localized at
${\x}_0$, is the time convolution of the Green's tensor with
${\J}_s$, integrated over the scattered volume,
\begin{equation}
\delta E^p({\x},t) \ =\ \int G^{pq}({\x}, t, {\x}_0, t_0) * \mu
{\dt {J_s^q}}({\x}_0,t_0) d V(\x_0),
\label{eq: E G J}
\end{equation}
with an implicit summation on the repeated indices and time
convolution represented by an asterisk. Therefore, the field $\delta
\E$ diffracted by a small perturbation in electromagnetic parameters
is calculated \cite{saintenoy98} from equation~(\ref{eq: E G J}),
using the secondary source terms expressed from equations~(\ref{eq:
res1}), (\ref{eq: contrasts 1}), (\ref{eq: contrasts 2}) and~(\ref{eq:
contrasts 3}), as
\begin{equation}
\begin{split}
\delta \E ({\x},t)
  & = \ \frac{1}{4 \pi r c^2} 
   \left[ A_{\eps}^* \left({\r} \times \dttwo \E 
    \left({\x}_0,t-\frac{r}{c}\right)\right) \times {\r} \right.  \\[3mm]   
  & +\ \mu {\sig} c^2 A_{\sig}^* \left({\r} \times \dt \E
    \left({\x}_0,t-\frac{r}{c}\right)\right) \times {\r}          \\[3mm]
  & \left. +\ A_{\mu}^* \left(\R_{inc} \times
    \dttwo \E  \left({\x}_0,t-\frac{r}{c}\right)\right) \times {\r} \right],
\label{eq: champ dif}
\end{split}
\end{equation}
where {\r} is the unit vector pointing from the diffracting point,
${\x}_0$, towards the observation point, \x. $\R_{inc}$ is the unit
vector in the direction of the incident wavefront displacement
(Figure~\ref{fig: champ incident}). The diffracted magnetic field
associated with the diffracted electric field is given by
equation~(\ref{eq: rot E}).

The analytical expression~(\ref{eq: champ dif}) of the diffracted
electric field allows for the separation of the contribution to the
total diffracted field of each type of anomaly. A point anomaly
diffracts an electric field $\delta \E$ that is the sum of three
terms. In each term of equation~(\ref{eq: champ dif}), the amplitude
of $\delta \E$ is proportional to the volumetric contrasts
$A^*_{\eps}$, $A^*_{\sig}$ or $A^*_\mu$, with a spatial dependence
(the cross products between {\r}, $\R_{inc}$, and the incident
electric field \E) and a time dependence (a first or second time
derivative of the incident electric field \E). The contribution to \E
produced by $A^*_{\eps}$ has the same spatial dependence as the
contribution produced by $A^*_{\sig}$, but does not have the same time
dependence. The distribution of the diffracted field amplitude,
normalized to its maximum value, over a sphere centered at the point
anomaly is called a radiation pattern. Thus, to have the same spatial
dependence implies the same radiation pattern. Taken separately, an
anomaly in dielectric permittivity has the same spatial radiation
pattern as an anomaly in electrical conductivity, but does not have
the same spatial radiation pattern as an anomaly in magnetic
permeability. Therefore, considering only two types of point
diffractors, an electric point diffractor and a magnetic point
diffractor, is justified if we are interested only in the spatial
radiation pattern and not in the time dependence of the signal.

It should be noted that the contribution to the total field of the
contrast in electrical conductivity is proportional to the electrical
conductivity of the surrounding medium, \sig. In our case where we
consider only low-conductive media, this contribution will be small.

\subsection{Radiation and polarization patterns}

Expression~(\ref{eq: champ dif}) shows that each type of diffracting
point can be described by a radiation pattern and a polarization
pattern (display of the diffracted electric field polarization and
amplitude over a sphere centered at the point anomaly). The radiation
pattern for an electric diffracting point (Figure~\ref{fig: radia
elec}a) is a torus centered at the point anomaly. No electric field
is diffracted in the direction of electric incident field (the axis of
the torus). The corresponding polarization pattern (Figure~\ref{fig:
radia elec}b) shows that the diffracted electric field is poloidal,
whereas the diffracted magnetic field is toroidal. These patterns are
the same as those obtained in the far-field for a small electric
dipole subjected to an incident electromagnetic field, where the axis
of the electric dipole is parallel to the incident electric field.

\begin{figure}
\includegraphics*[width=7cm]{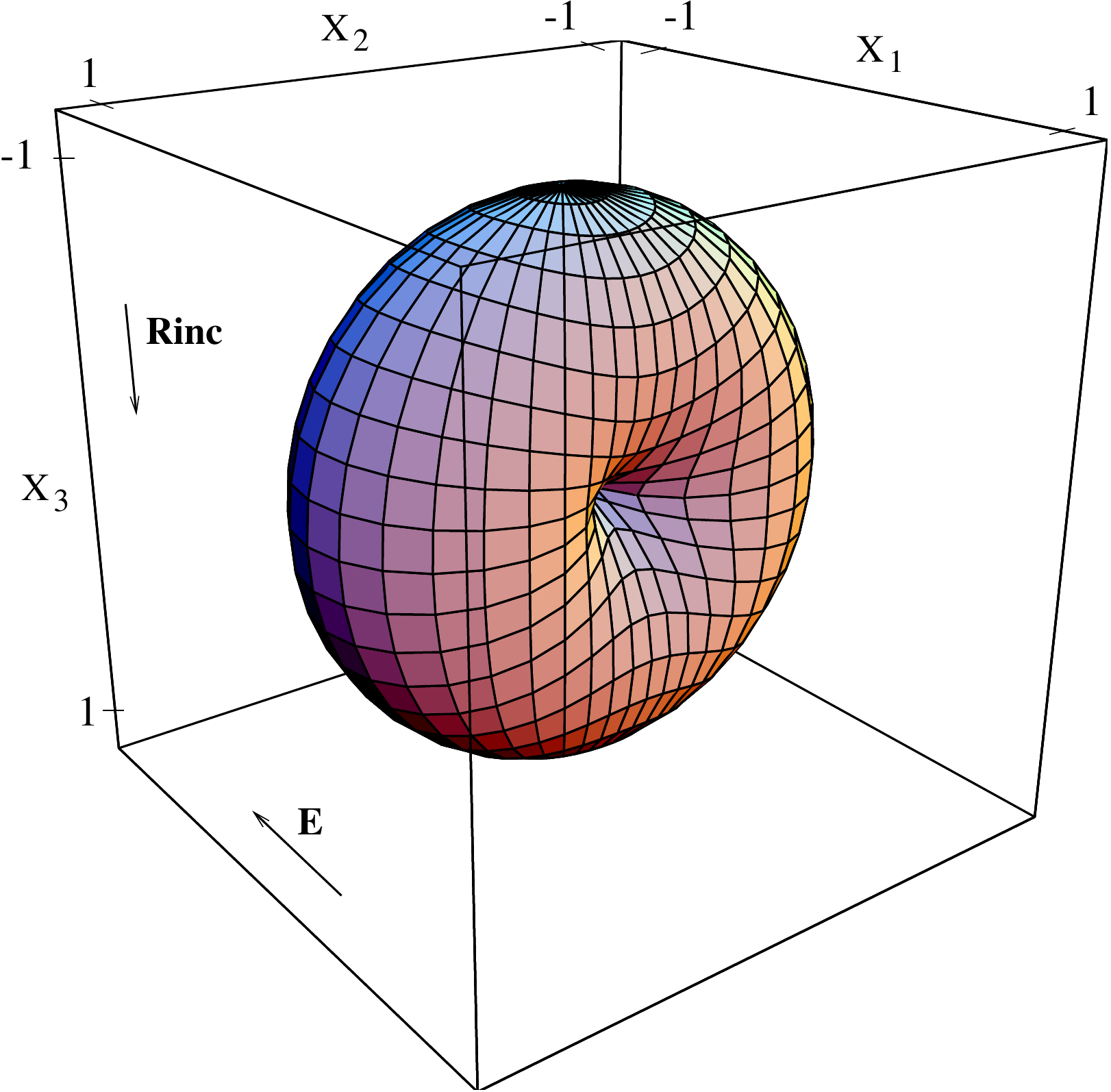}
\hfill
\includegraphics*[width=7cm]{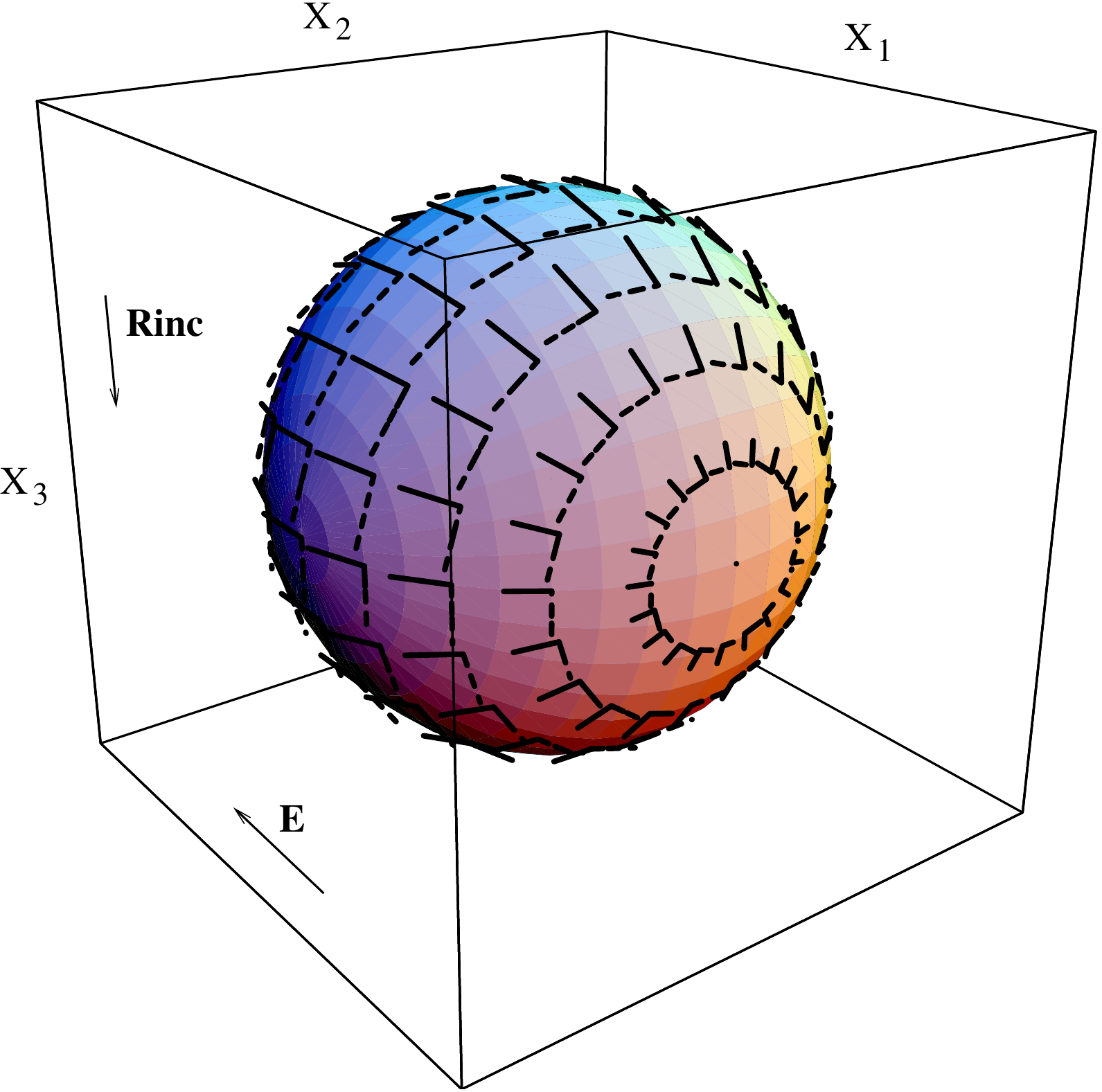}

\hspace{3.5cm} a) \hspace{7cm} b)
\vspace{5mm}

\caption{(a) Radiation and (b) polarization patterns of a positive
electric point anomaly when illuminated by an electromagnetic wave
moving in the $\protect\R_{inc}$ direction with an electric field \E
in the $X_1$ direction. In (b), the diffracted electric field (solid
lines) is poloidal whereas the corresponding magnetic field (dashed
lines) is toroidal.}

\label{fig: radia elec}
\end{figure}

\begin{figure}
\includegraphics*[width=7cm]{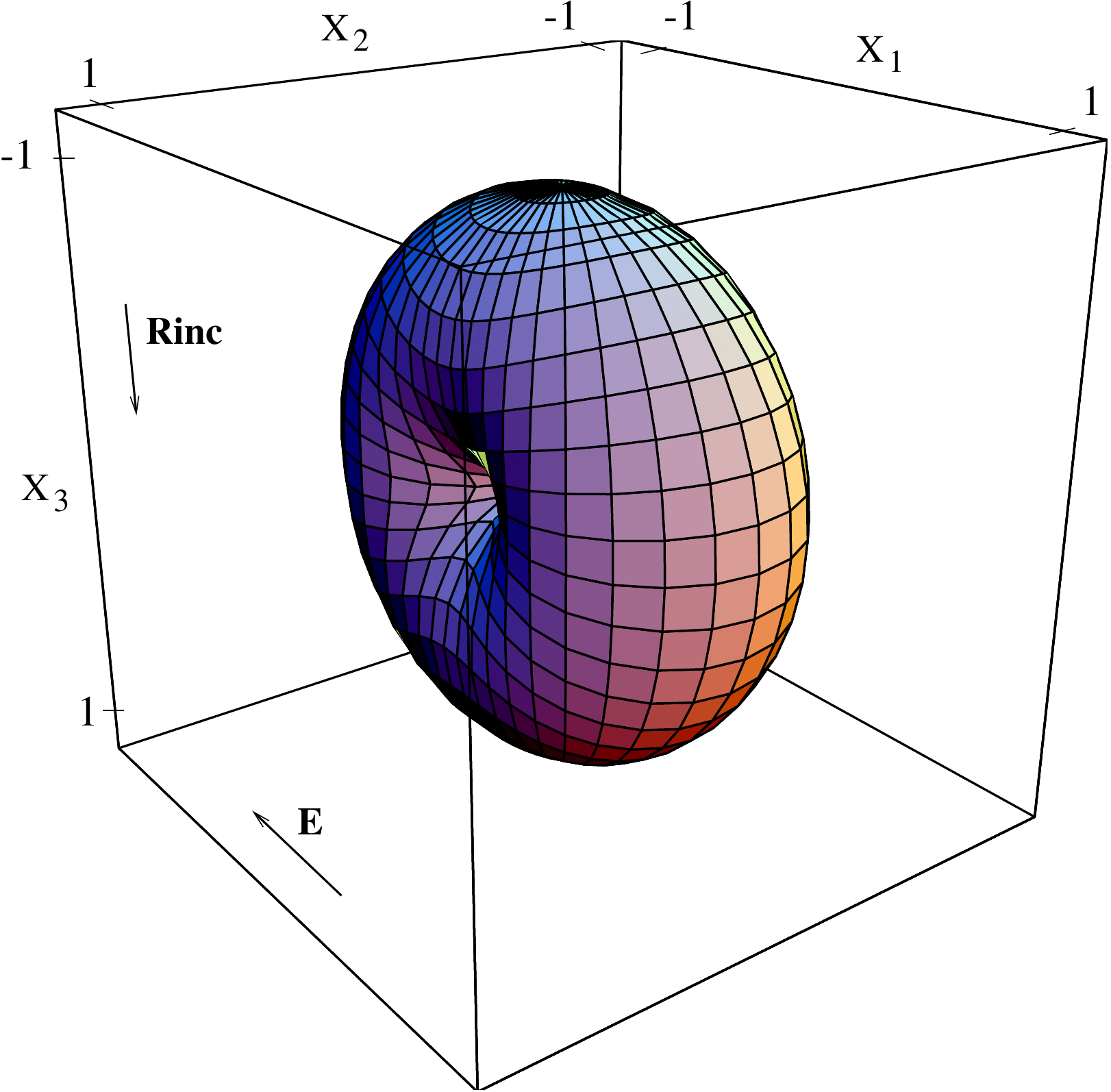}
\hfill
\includegraphics*[width=7cm]{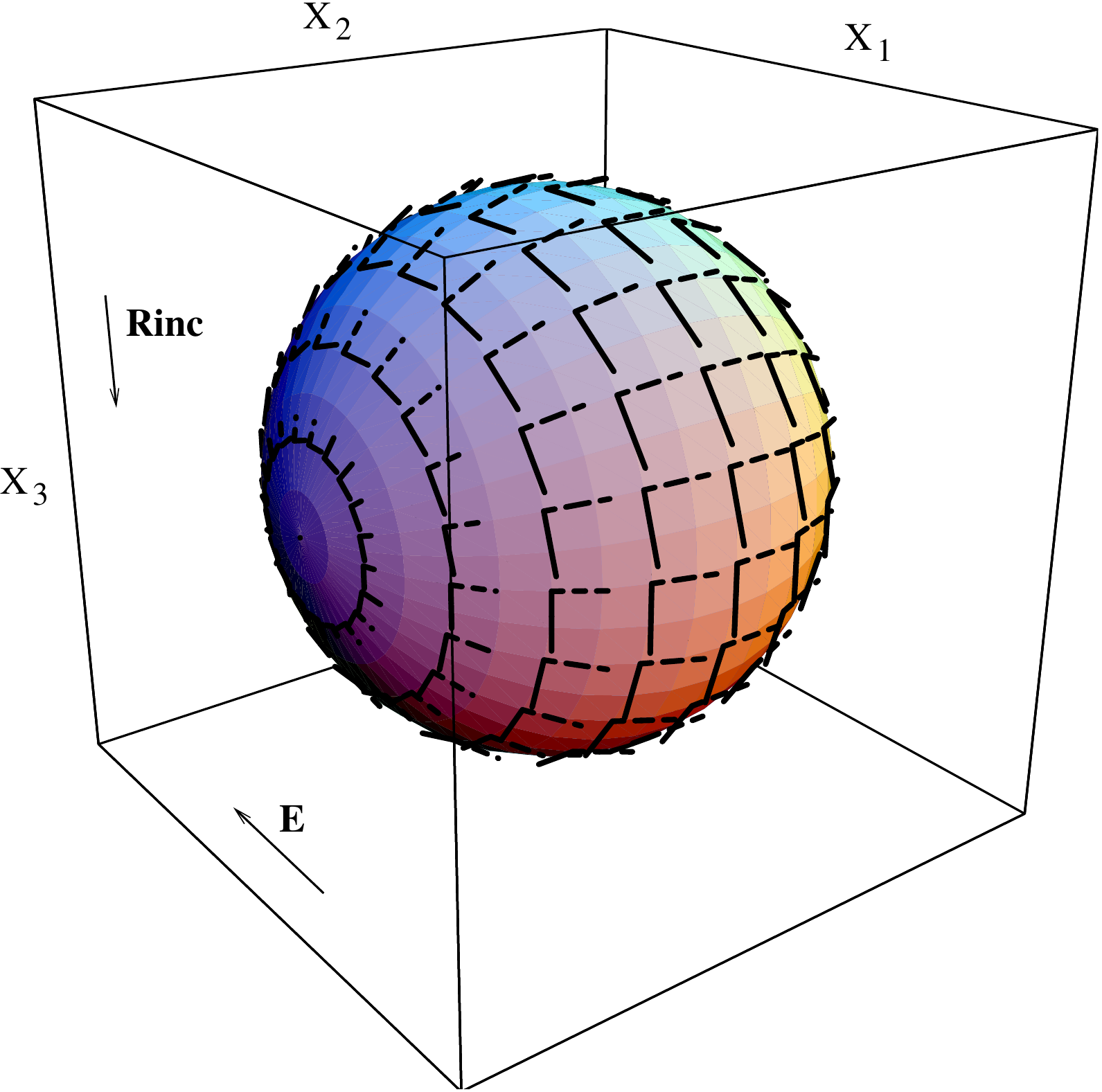}

\hspace{3.5cm} a) \hspace{7cm} b)
\vspace{5mm}

\caption{(a) Radiation and (b) polarization patterns of a positive
magnetic point anomaly when illuminated by an electromagnetic wave
moving in the $\protect\R_{inc}$ direction with an electric field \E
in the $X_1$ direction. In (b), the diffracted electric field (solid
lines) is toroidal whereas the corresponding magnetic field (dashed
lines) is poloidal.}

\label{fig: radia magnet}
\end{figure}

A magnetic diffracting point behaves, in the far-field, like a small
magnetic dipole when it is subjected to an electromagnetic field. The
diffracted magnetic field is the same that of the electric field
diffracted by a small electric dipole parallel to the incident
magnetic field. Figure~\ref{fig: radia magnet}a displays the
magnetic dipole radiation pattern, which is a torus perpendicular to
that in Figure~\ref{fig: radia elec}a. Figure~\ref{fig: radia
magnet}b shows the corresponding polarization pattern. The magnetic
field diffracted by a magnetic diffracting point is poloidal, whereas
the diffracted electric field is toroidal.

\subsection{Point diffractor in (${\eps}_e$, $\mu$) and ($Z$, $c$)
parameterizations}

A point contrast in electrical conductivity diffracts an
electromagnetic field with the same radiation pattern as that of a
point contrast in dielectric permittivity. However the field
diffracted by the conductivity contrast alone depends on the
first-time derivative of the incident electric field, whereas, the
field diffracted by the permit\-ti\-vi\-ty contrast alone depends on
the second-time derivative of the electric incident field [equation
(\ref{eq: champ dif})]. Therefore, dielectric permittivity and
conductivity can be merged, introducing a fictitious time dependence of
the electric parameter, into the effective dielectric permittivity
${\eps}_e$,
\begin{equation}
{\eps}_e \delta(t) ={\eps} \delta(t) + {\sig} {\rm H}(t),
\end{equation}
where ${\rm H}(t)$ is the Heaviside function. In the Fourier domain,
considering an harmonic dependent electromagnetic field,
the effective dielectric permittivity is
\begin{equation}
{\eps}_e\ =\ {\eps} + \frac{\sig}{\i \omega},
\end{equation}
where $\omega$ is the incident field frequency \cite{jackson98}. With
factorization of this parameter, equation~(\ref{eq: rot H}) becomes
\begin{equation}
{\rot} \left( \frac{\B ({\x},t)}{\mu ({\x})} \right)\ -\ {\eps}_e ({\x}) * \dt
\E ({\x},t)\ =\ {\J}_s ({\x},t).
\end{equation}

Following the above approach, a small perturbation in logarithmic
effective permittivity, subjected to an incident electromagnetic
field, acts as a secondary source of current density,
\begin{equation}
{\J}_s ({\x},t)\ =\ {\eps}_e ({\x}) \delta {\eps}_e^* ({\x}) * \dt \E ({\x},t).
\end{equation}
Consequently, a point anomaly that is described by
\begin{equation}
\delta {\eps}_e^*({\x})\ =\ A_{ef\!f}^*\ \delta_{1/V}({\x} - {\x}_0),
\label{eq: epseff}
\end{equation}
and
\begin{equation}
\delta \mu^*({\x})\ =\ A_{\mu}^*\ \delta_{1/V}({\x} - {\x}_0),
\label{eq: muagain}
\end{equation}
in an isotropic, homogeneous, linear, low-conductivity
medium diffracts an electric field \cite{saintenoy98}
\begin{equation}
\begin{split}
\delta \E ({\x},t)\ = \ \frac{\eps \mu}{4 \pi r}
  & \left[ A_{ef\!f}^* * \left({\r} \times
     \dttwo \E \left({\x}_0,t-\frac{r}{c}\right)\right)
      \times {\r} \right.\\       
  & \left. +\ A_{\mu}^* * \left({\R}_{inc} \times
     \dttwo \E  \left({\x}_0,t-\frac{r}{c}\right)\right) \times {\r} \right].
\label{eq: champ dif bis}
\end{split}
\end{equation}
Terms $A_{ef\!f}^*$ and $A_{\mu}^*$ are the perturbations in
logarithmic parameters $\eps_{ef\!f}^*$ and $\mu^*$ multiplied by the
volume of the anomaly.

\begin{figure}
\includegraphics*[width=7cm]{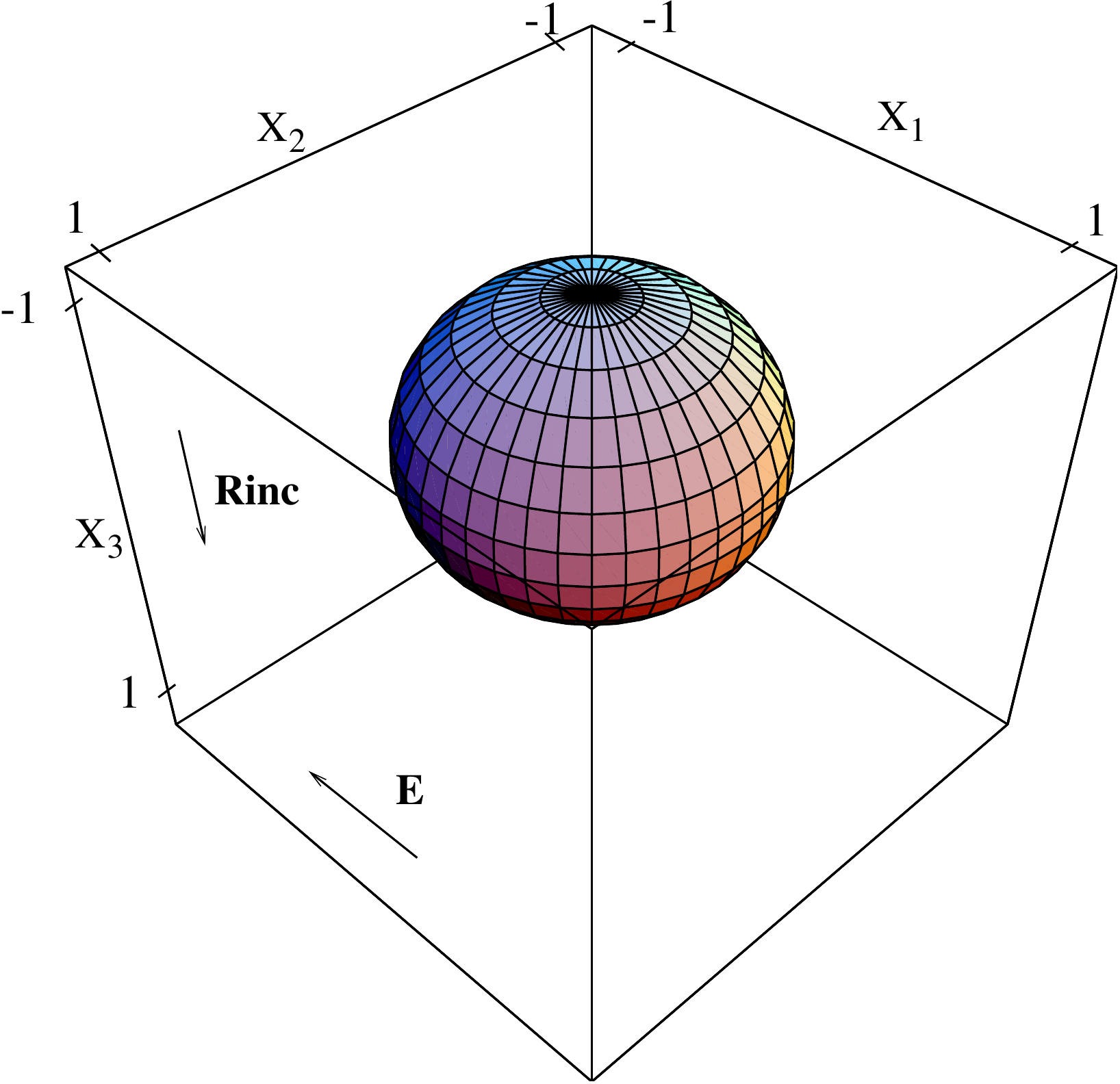}
\hfill
\includegraphics*[width=7cm]{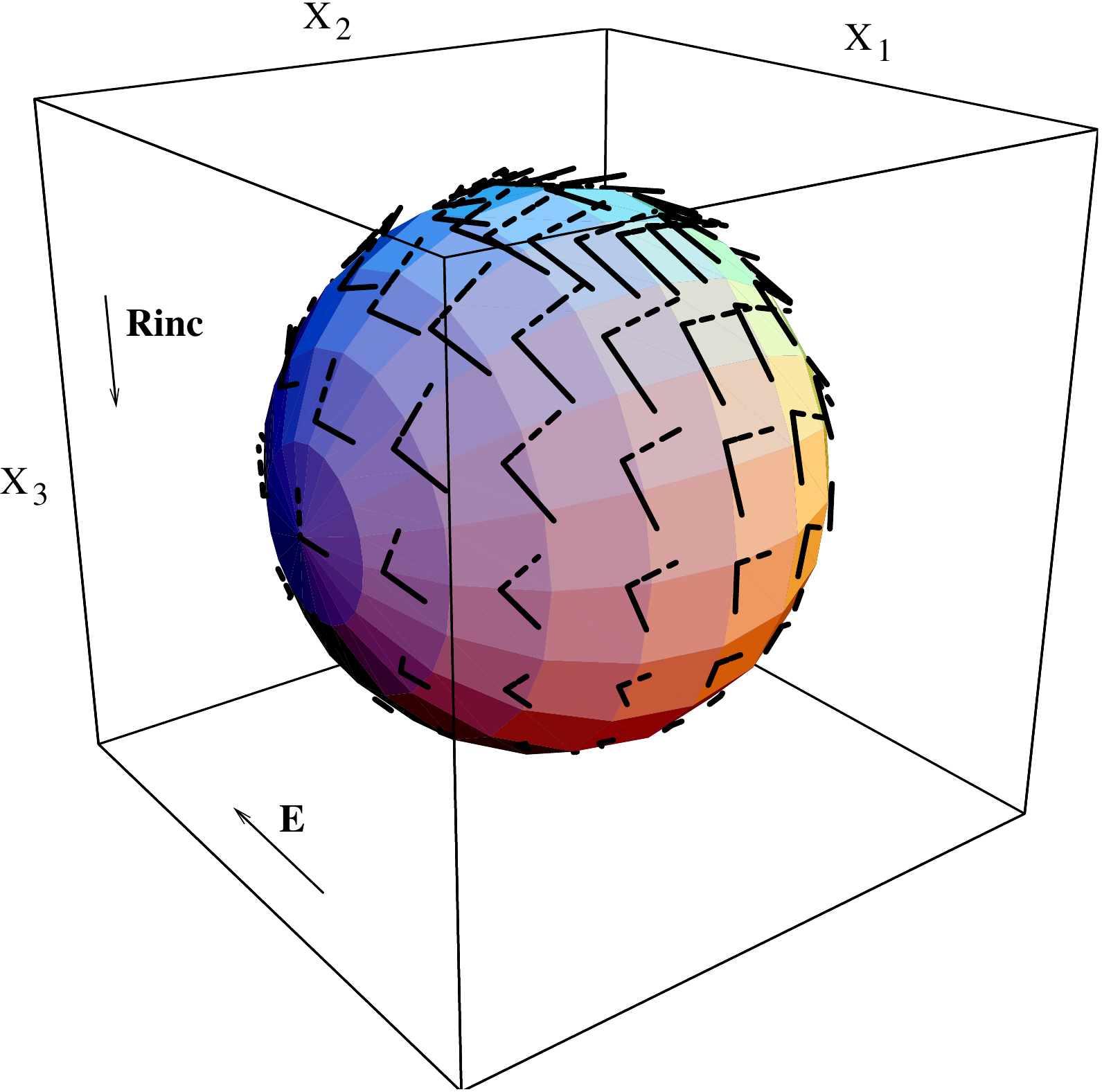}

\hspace{3.5cm} a) \hspace{7cm} b)
\vspace{5mm}

\caption{(a) Radiation and (b) polarization patterns of a positive
  impedance point anomaly when illuminated by an electromagnetic wave
  moving in the $\protect\R_{inc}$ direction with an electric field \E
  in the $X_1$ direction. In (b), the diffracted electric field is
  represented by solid lines, and its associated magnetic field by
  dashed lines, on the sphere.}

\label{fig: radia z}
\end{figure}

\begin{figure}
\includegraphics*[width=7cm]{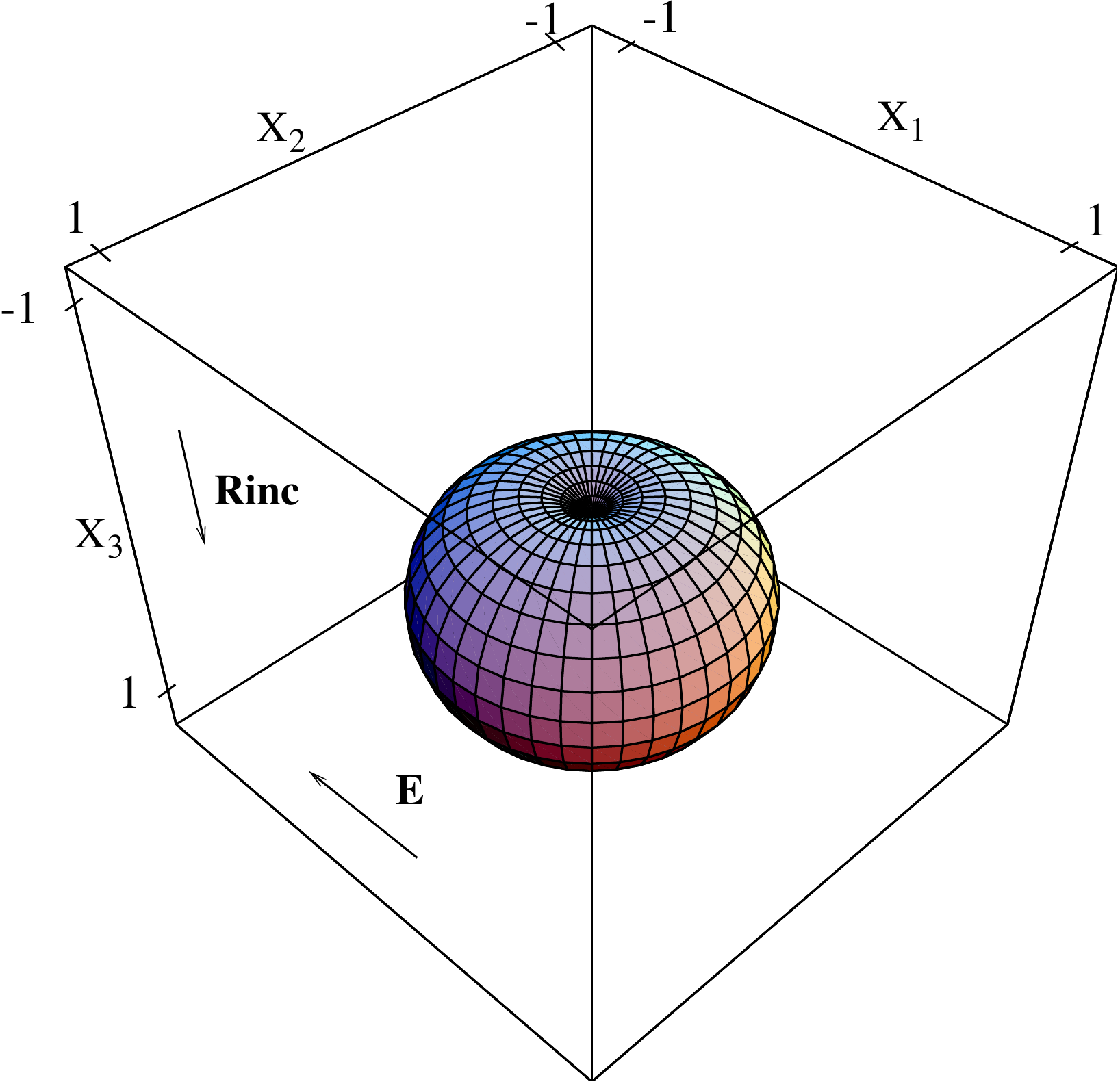}
\hfill
\includegraphics*[width=7cm]{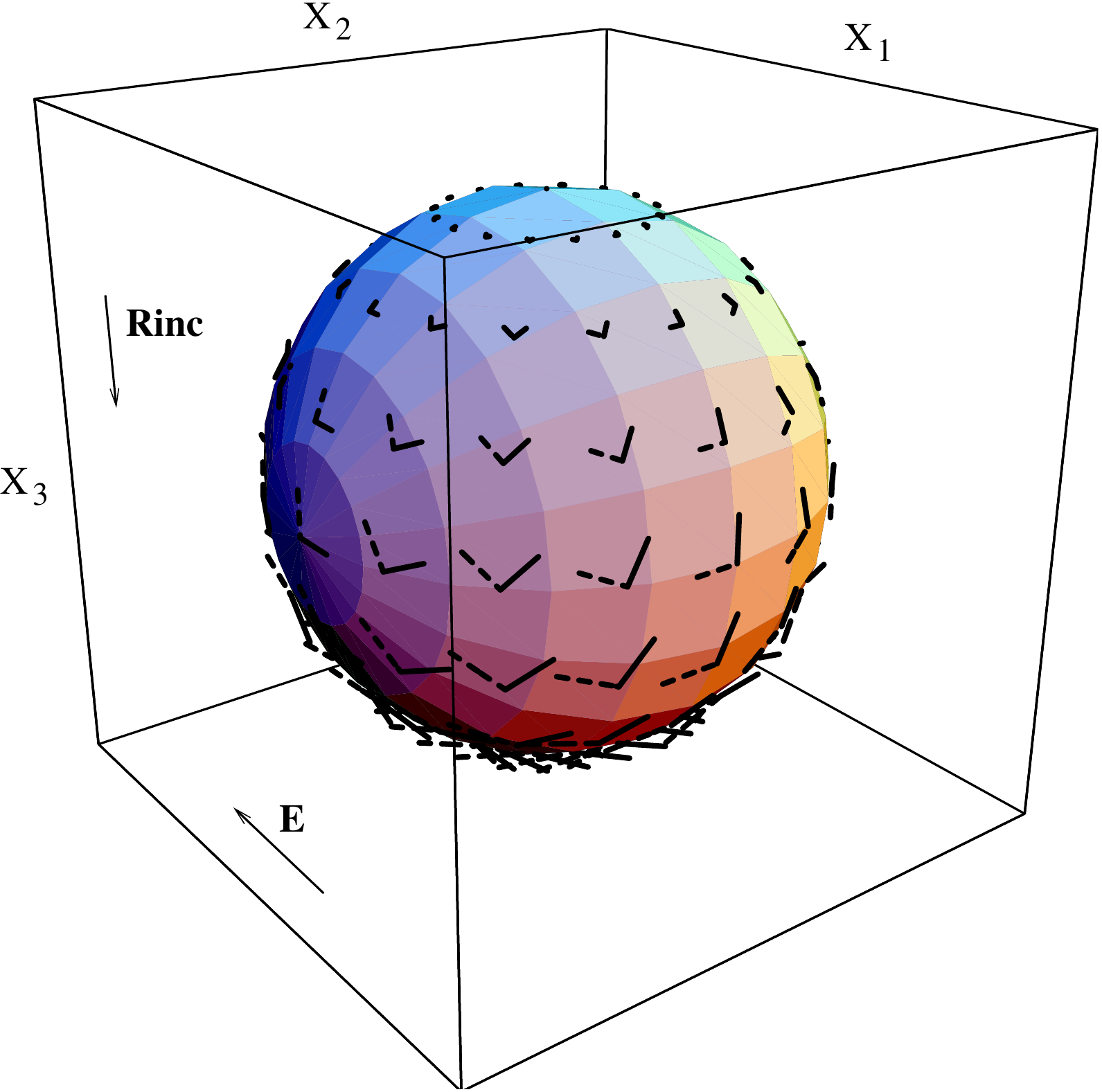}

\hspace{3.5cm} a) \hspace{7cm} b) 
\vspace{5mm}

\caption{(a) Radiation and (b) polarization of a positive velocity
  anomaly when illuminated by an electromagnetic wave moving in the
  $\protect\R_{inc}$ direction with an electric field \E in the $X_1$
  direction. In (b), the diffracted electric field is represented by
  solid lines, and its associated magnetic field by dashed lines, on
  the sphere.}

\label{fig: radia c}
\end{figure}

Another parameterization must be considered as well. The effective impedance
$Z$ and the effective velocity $c$ defined from the magnetic
permeability and the effective dielectric permittivity as
\begin{equation}
Z=\sqrt{\frac{\mu}{{\eps}_e}}
\label{eq: Zeff}
\end{equation}
and
\begin{equation}
c=\sqrt{\frac{1}{\mu {\eps}_e}}.
\label{eq: ceff}
\end{equation}
Then, combining~(\ref{eq: epseff}) and~(\ref{eq: muagain})
with~(\ref{eq: Zeff}) and~(\ref{eq: ceff}) results in
\begin{equation}
A_Z^*\ =\ \frac{1}{2}(A_\mu^* - A_{ef\!f}^*)
\end{equation}
and
\begin{equation}
A_c^*\ =\ \frac{1}{2}(-A_\mu^* - A_{ef\!f}^*).
\end{equation}
Terms $A_{Z}^*$ and $A_{c}^*$ are the perturbations in logarithmic
parameters $Z^*$ and $c^*$ multiplied by the volume of the anomaly.
Using $A_Z^*$ and $A_c^*$, equation~(\ref{eq: champ dif bis}) becomes
\cite{saintenoy98}
\begin{equation}
\begin{split}
\delta \E ({\x},t)\ = \ \frac{{\eps} \mu}{4 \pi r} 
  & \left[ A_{Z}^* * \left((-{\r} + {\R}_{inc}) \times
     \dttwo \E \left({\x}_0,t-\frac{r}{c}\right)\right)
      \times {\r} \right. \\[3mm] 
  & \left. -\ A_{c}^* * \left(({\r} + {\R}_{inc}) \times
     \dttwo \E  \left({\x}_0,t-\frac{r}{c}\right)\right) \times {\r} \right].
\label{eq: champ dif ter}
\end{split}
\end{equation}
The radiation and polarization patterns defined by equation~(\ref{eq:
champ dif ter}), associated with a point diffractor described by the
parameters $Z$ and $c$, are displayed in Figures~\ref{fig: radia z}
and~\ref{fig: radia c}.

\section{DISCUSSION}

\subsection{Validity of the diffracted field analytical expressions}

The derivation of expressions~(\ref{eq: champ dif}), (\ref{eq: champ
  dif bis}) and~(\ref{eq: champ dif ter}), required many assumptions
in addition to isotropy and linearity. First, to simplify the
expression of the Green's tensor and to use it in the time domain,
only the far-field term was considered,
\begin{equation}
\lambda\ \ll\ r,
\label{eq: assumption 1}
\end{equation}
and, the conductivity of the surrounding medium was assumed to be
small,
\begin{equation}
\sig\ \ll\ \eps \omega.
\label{eq: assumption 2}
\end{equation}
To model some actual \GPR data, the computation should be done in the
frequency domain and the complete expression of the Green's
tensor~(\ref{eq: green frequence}) should be used. However, to get a
diagnostic characterization, of the effects of anomalies in electrical
and magnetic properties in a homogeneous medium, on the diffracted
electric field, assumption~(\ref{eq: assumption 2}) is not limiting.
Indeed, an electrical conductivity anomaly has the same radiation
pattern as a dielectric permittivity anomaly. However, because of
assumption~(\ref{eq: assumption 1}) our diagnostic will be correct
only for far-field data.

An approach for analyzing scattering effects in a dispersive medium is
discussed in Saintenoy \shortcite{saintenoy98}. In this paper, the relative
importance of electrical and magnetic effects are discussed for a
single frequency. Only non-dispersive media are considered. The
frequency dependence will be the subject of future work and is not
discussed here.

A less obvious aspect should be noted about our computation of
equation~\ref{eq: champ dif}. We wish to perform the modeling of the
wavefield using a first order approximation (it happens that ``first
order (Taylor) approximation'' and ``first order Born approximation''
are equivalent). We linearize equations~(\ref{eq: var linearisees})
and keep only first order terms in equation~(\ref{eq: res1}). To
actually use an approximation like $\exp [\delta m^*(\x)]\ =\ 1\ +\ 
\delta m^*(\x)$, $\delta m^*(\x)$ has to be small in some sense. A
sensible measurement of the smallness of a function is that the
$L_2$-norm of the function,
\begin{equation}
||\delta m^*(\x)||_2\ =\ V |\delta m^*|,
\end{equation}
must be small. Practically, in our context, the linearization is valid
when 
\begin{equation}
|A_m^*| \ll\ \lambda^3,
\label{eq: assumption 3}
\end{equation}
with m being \eps, \sig and $\mu$ successively. If the logarithmic
parameter contrast goes to infinity, the perturbation volume will have
to go to zero in order for the Born approximation to be true. This
criterion is consistent with several papers
\cite{dehoop91,gritto95,hudson81} where the domain of validity of the
Born approximation is discussed.

\subsection{Importance of each parameter in the ({\eps}, {\sig}, $\mu$) set}

Expression~(\ref{eq: champ dif}) of the diffracted field allows
for the comparison of the relative importance of contrasts in the
three parameters \eps, \sig and $\mu$. For a monochromatic
incident electric field \E, the maximum amplitude of the
field diffracted by a dielectric permittivity anomaly is
\begin{equation}
|\delta \E|_{\eps}\ =\ \frac{1}{4 \pi r c^2} A_{\eps}^* \omega^2 |\E|,
\end{equation}
by an electrical conductivity anomaly,
\begin{equation}
|\delta \E|_{\sig}\ =\ \frac{1}{4 \pi r} \mu {\sig} A_{\sig}^* \omega
|\E|,
\end{equation}
and by magnetic permeability anomaly,
\begin{equation}
|\delta \E|_\mu\ =\ \frac{1}{4 \pi r c^2} A_{\mu}^* \omega^2 |\E|,
\end{equation}
with $\omega$ as the frequency of the incident wave. The
contributions of \eps, \sig, $\mu$ are then proportional to ${\eps}
A_{\eps}^* \omega$, ${\sig} A_{\sig}^*$ and ${\eps} A_\mu^* \omega$,
respectively.

Consider a small sphere composed of a mixture of 20\% iron filings and
80\% silica sand matrix, in a surrounding homogeneous dry sand,
subjected to an electromagnetic field of frequency 600 MHz. Assume
the sphere is small enough and deep enough to satisfy
assumptions~(\ref{eq: assumption 1}) and (\ref{eq: assumption 3}).
Realistic parameter values for the mixture of iron filings and silica
sand matrix (${\eps}_a, {\sig}_a, \mu_a$) and the dry sand (${\eps},
{\sig}, \mu$) can be found in Olhoeft and Capron
\shortcite{olhoeft93},
\begin{alignat*}{2}
{\eps}    &= 2.2\ {\eps}_0,                    & \qquad
{\eps}_{a}&= 2.8\ {\eps}_0,\\
\mu     &= \mu_0,                       & \qquad
\mu_{a} &= 1.2\ \mu_0,\\
{\sig}    &= 4.6\ 10^{-5}\ {\rm S}/{\rm m},    & \qquad
{\sig}_{a}&= 6.92\ 10^{-5}\ {\rm S}/{\rm m}.
\end{alignat*}
Since
\begin{equation}
\eps \omega \approx 0.01 S/m,
\end{equation}
we verify that $\sig \ll \eps \omega$. 
Then,
$$
{\eps} \ln \frac{{\eps}_a}{\eps} \omega \approx  2.8\,10^{-3}\ \text{\rm S/m}
$$
for the dielectric permittivity contrast contribution,
$$
{\sig} \ln \frac{{\sig}_a}{\sig} \approx  1.9\,10^{-5}\ \text{\rm S/m}
$$
for the electrical conductivity contrast contribution, and
$$
{\eps} \ln \frac{\mu_a}{\mu} \omega \approx  2.2\,10^{-3} \text{\rm S/m}
$$
for the magnetic permeability contrast contribution.

In this example, the electric conductivity contrast contribution is
negligible compared to that of the two other contributions. Terms in
front of the dielectric permittivity and magnetic permeability
contrasts are proportional to $\omega$ while the electrical
conductivity contrast contribution does not depend on $\omega$, and
the surrounding medium has low conductivity. However, this example was
chosen to highlight that the contribution of the permeability contrast
to the amplitude of the total diffracted field can not be neglected
{\em a priori}. L\'azaro-Mancilla {\em et al.} \shortcite{lazaro96}
find the same result but think that the magnetic permeability does not
vary significantly in most earth materials. On the other hand, ferrous
metals, magnetite, hematite have relative permeability values that
differ significantly from 1. Olhoeft and Capron's report
\shortcite{olhoeft93} contains measurements of several natural sands
with relative magnetic permeabilities higher than 1.2. Olhoeft
\shortcite{olhoeft98} emphasizes that the materials found on Mars have
relative magnetic permeability significantly greater than 1 as do
sampled soils coming from Arizona, Idaho, Colorado, Hawaii, Australia
and Canada (Olhoeft, personal communication). Relative magnetic
permeability higher than 1 is more common than people would like to
think. Therefore we cannot neglect the magnetic permeability effect on
radar data without an explicit justification.

\subsection{New geometry of acquisition ?}

\begin{figure}
\centerline{\includegraphics*[width=12cm]{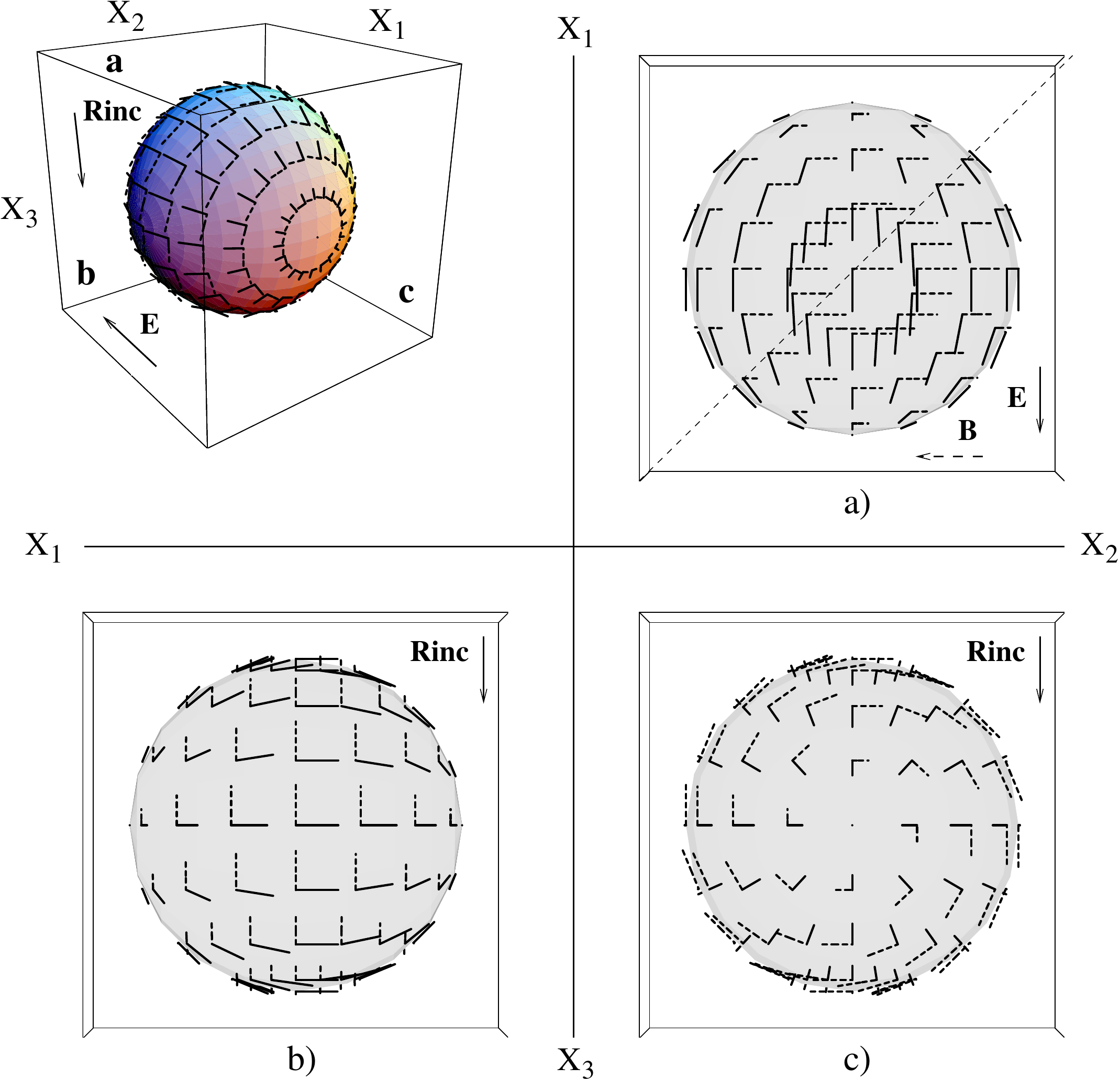}}
\vspace{5mm}

\caption{To visualize the effects of different acquisition geometries
  on \GPR data due to an electric point diffractor, Figure~\ref{fig:
    radia elec}b is redisplayed here (upper left) and viewed along the
  three axis; (a) along the $X_3$ axis; (b) along the $X_2$ axis; (c)
  along the $X_1$ axis.}

\label{fig: dip elec 3 vues}
\end{figure}

\begin{figure}
\centerline{\includegraphics*[width=12cm]{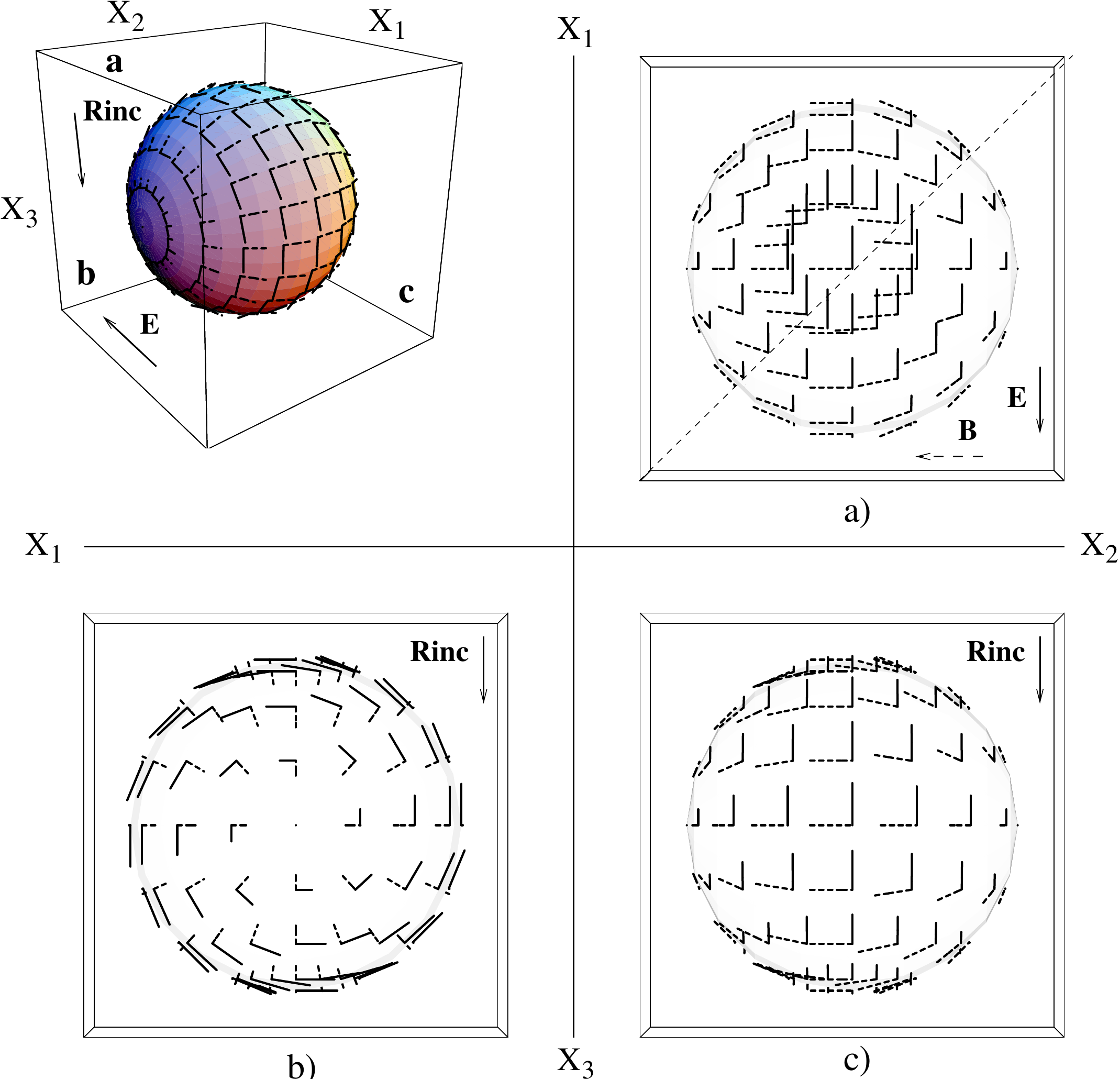}}
\vspace{5mm}

\caption{To visualize the effects of different acquisition geometries
  on \GPR data due to a magnetic point diffractor, Figure~\ref{fig:
    radia elec}b is redisplayed here (upper left) and viewed along the
  three axis; (a) along the $X_3$ axis; (b) along the $X_2$ axis; (c)
  along the $X_1$ axis.}

\label{fig: dip mag 3 vues}
\end{figure}

From equation~(\ref{eq: champ dif}), the radiation and polarization
patterns associated with contrasts in each parameter are displayed in
Figures~\ref{fig: radia elec} and \ref{fig: radia magnet}.
Figures~\ref{fig: dip elec 3 vues} and~\ref{fig: dip mag 3 vues}
summarize the same information on radiation and polarization, showing
the effects of different acquisition geometries. For example, the
electromagnetic field diffracted by a point anomaly recorded with
zero-offset measurements is shown in the middle of Figures~\ref{fig:
dip elec 3 vues}a and~\ref{fig: dip mag 3 vues}a. The diffracted
field in both cases has the same polarization direction. Magnetic and
electric effects are then indistinct in a zero-offset \GPR experiment.

Acquisition using two radar antennas parallel to each other and
perpendicular to the acquisition profile consists of recording the
diffracted field along the $X_2$ axis. Along this axis, the amplitude
of the diffracted field is constant with the offset in Figure~\ref{fig:
dip elec 3 vues}a, whereas it is decreasing with the offset in
Figure~\ref{fig: dip mag 3 vues}a. Therefore, when data are
recorded with both antennas $90^{\circ}$ to the plane of acquisition, the
dielectric behavior differs from the magnetic behavior. For a magnetic
anomaly, the amplitude of the diffracted field depends on the offset;
this is not true for a dielectric anomaly.

When using two radar antennas parallel to each other and parallel to
the acquisition profile, the field diffracted by an anomaly is
recorded along the $X_1$ axis. Along this axis, the amplitude of the
diffracted field decreases with the offset in Figure~\ref{fig: dip
  elec 3 vues}a, whereas it is constant with the offset in
Figure~\ref{fig: dip mag 3 vues}a (because of the deformation due to
the projection on a plane, it might be easier to look at the 3D
radiation pattern in this case). Thus, when recording data with both
antennas parallel to the plane of acquisition, the amplitude depends
on the offset for a dielectric anomaly, and for a magnetic anomaly it
does not.

Another interesting method for acquisition of data would be to place
the source antenna at $45^{\circ}$ from the profile direction. The
result of this experiment would be seen along an axis that comes in at
an angle of $45^{\circ}$ to the $X_1$ and $X_2$ axes, in
Figures~\ref{fig: dip elec 3 vues}a and~\ref{fig: dip mag 3 vues}a.
The electric field diffracted by a dielectric point anomaly would not
be in the same direction as that diffracted by a magnetic point
anomaly. Consequently, multi-component measurements can, in this case,
help to distinguish magnetic effects from dielectric effects.
Unfortunately, the 3 dB beam width of a finite-size resistively loaded
horizontal electric dipole lying on a low-loss dielectric half-space
is roughly $50^{\circ}$ \cite{arcone95}. Thus, our theoretical
observations might be hampered in real \GPR experiments by the wide
radiation pattern of the antennas.

\subsection{The inversion parameters: Z and c}

Because of the difficulty in separating magnetic and dielectric effects, Z
and c will be the keys to the inverse problem. Indeed,
Figures~\ref{fig: radia z} and~\ref{fig: radia c} show that
zero-offset data are governed by the variations in the electromagnetic
impedance alone, whereas large-offset data contain information on the
effective velocity. To illustrate this behavior, reconsider the 5 cm
sphere composed of a mixture of 20\% iron fillings and 80\% of silica
sand matrix, at 4 meter depth in a homogeneous dry sand, and imagine a
Common Midpoint \GPR experiment centered above the sphere with
antennas centered at 600 MHz.

The source antenna is emitting an electric field \E that behaves with
respect to time as a second order Ricker function,
\begin{equation}
\E(\x_0,t)=- \frac{c}{4 \pi k r_0} (1 -2 \pi k^2 (t -
\frac{r_0}{c})^2 ) \exp [-\pi k^2 (t - \frac{r_0}{c})^2],
\label{eq: ricker}
\end{equation}
where $c$ is the wave speed in the dry sand, $k$ equals $600$ MHz, and
$r_0$ is the distance between the source antenna and the sphere; this
distance depends on the offset between the two radar antennas. The
antenna radiation pattern is that of a small dipole at an
interface between air and the host medium, for which equations are
given by Engheta \shortcite{engheta82}.

Realistic parameter values for the mixture of iron fillings and silica
sand matrix (${\eps}_a, {\sig}_a, \mu_a$) and dry sand (${\eps},
{\sig}, \mu$) have been given in the preceding numerical example. It
follows that the speed of the propagating wave in dry sand is
\begin{equation}
c \approx \sqrt{\frac{1}{\mu_0\,2.2\,\eps_0}} \approx 0.2\ {\rm m}/{\rm
ns},
\end{equation}
its main wavelength,
\begin{equation}
\lambda \approx \frac{c}{k} \approx 0.33\ {\text m},
\end{equation}
and, the product of dry sand dielectric constant and the dominant frequency,
\begin{equation}
\eps \omega \approx 1.1\ 10^{-2}\ {\text S/m}.
\end{equation}
Those numerical values satisfy conditions~(\ref{eq: assumption 1}),
(\ref{eq: assumption 2}), and~(\ref{eq: assumption 3}).

\begin{figure}
\begin{center}
\includegraphics*[width=7.5cm]{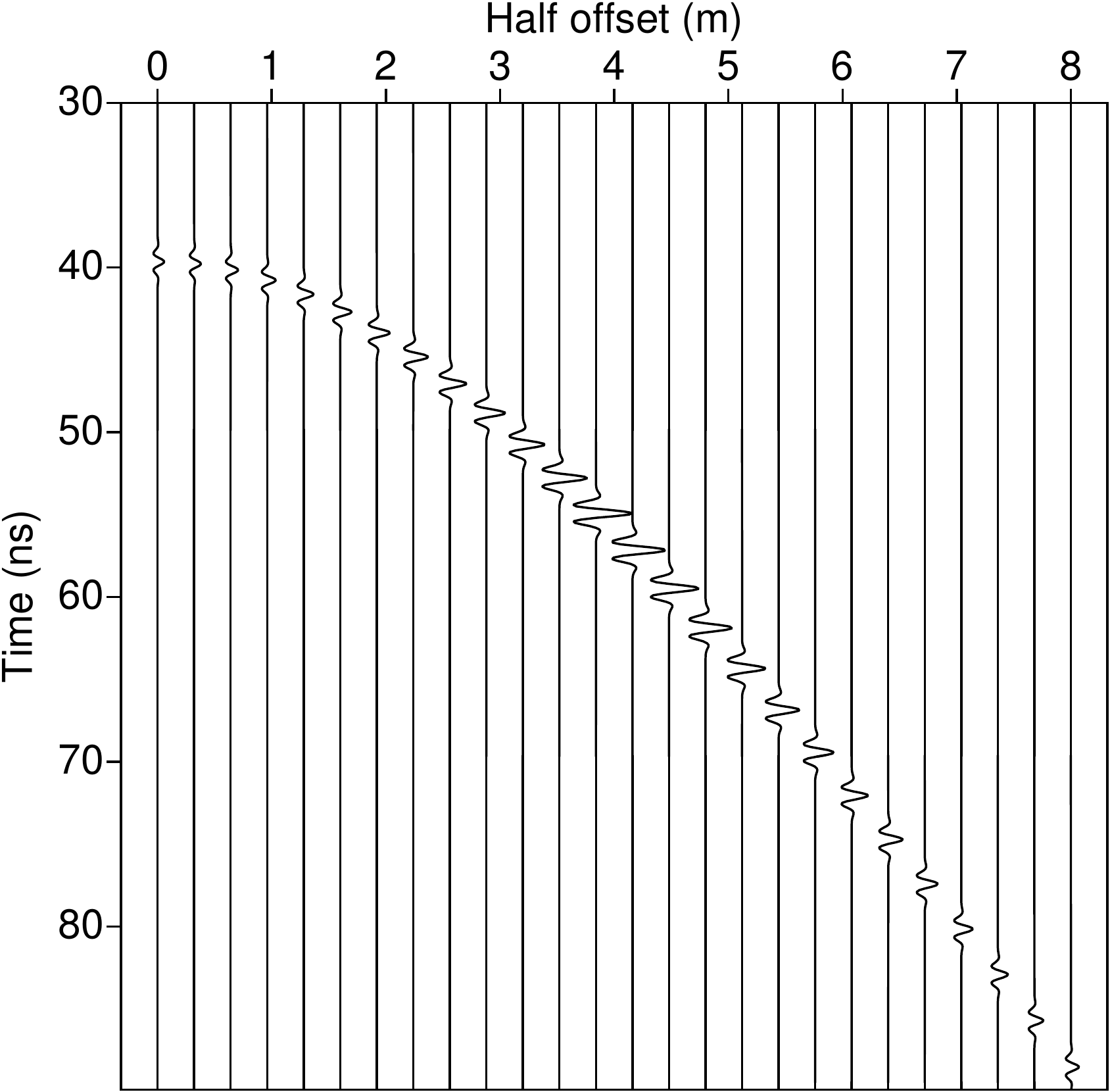}\\
\centerline{a)}
\vspace{5mm}
\includegraphics*[width=7.5cm]{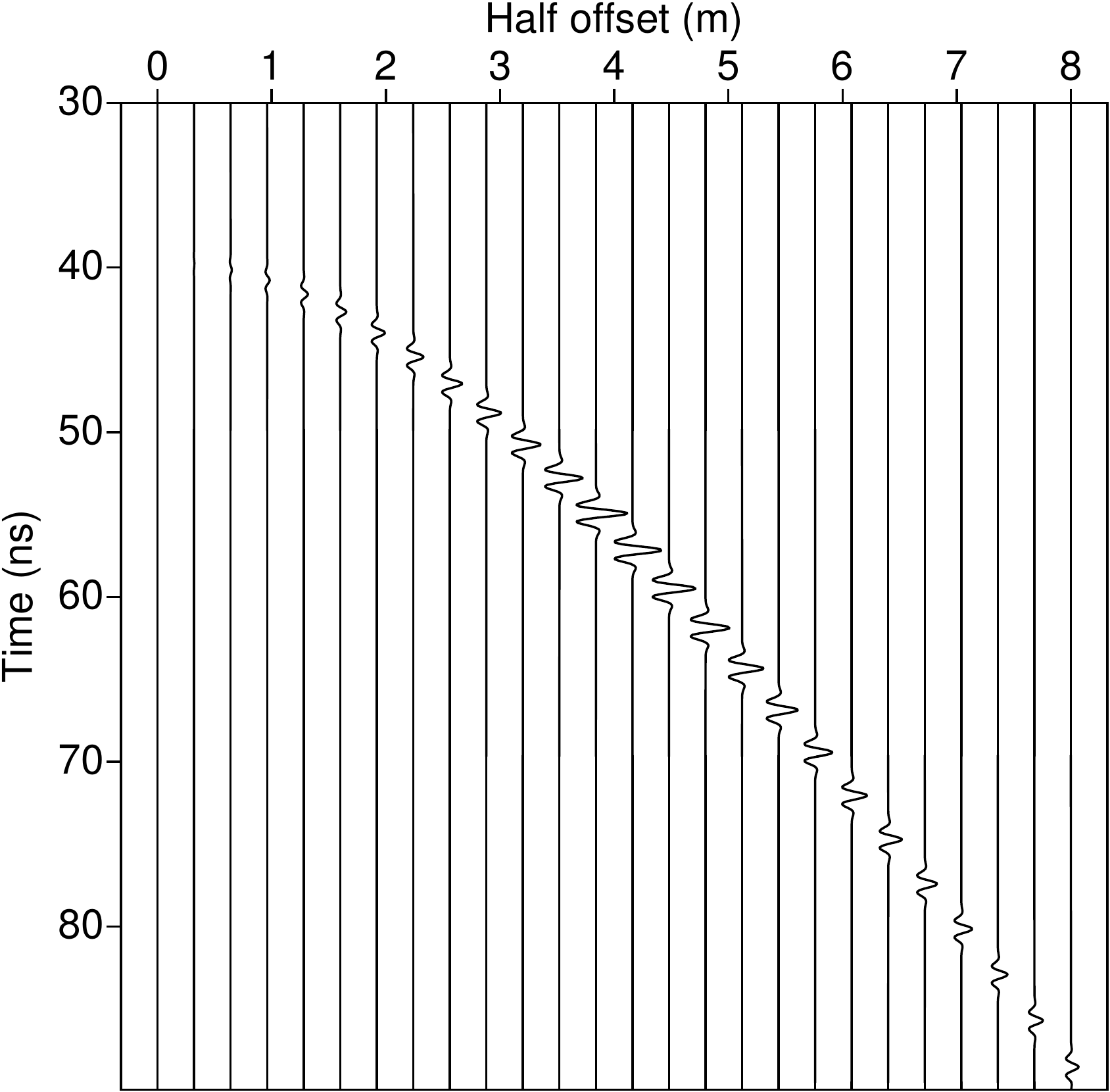}
\hfill
\includegraphics*[width=7.5cm]{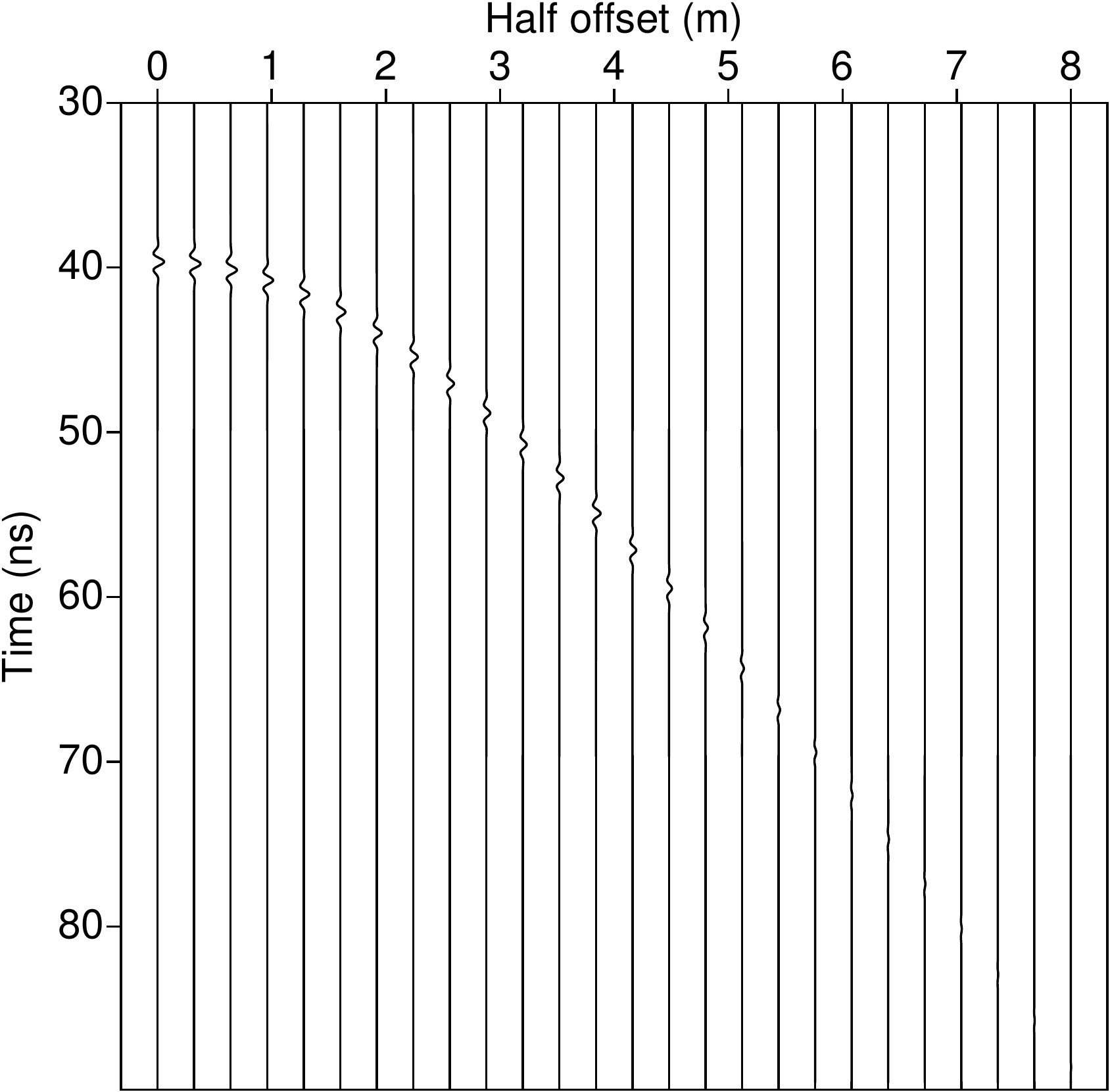}\\
\end{center}
\hspace{3.5cm} b) \hspace{7cm} c)
\vspace{10mm}

\caption{(a) Theoretical radargram for a common midpoint acquisition
over a small sphere composed of a mixture of 20\% iron fillings and
80\% of silica sand matrix at 4 meter depth in dry sand. The source
has the same radiation pattern as that of a small dipole and a second
order Ricker time dependence, centered at 600 MHz. The receiver has
the same radiation pattern as the source. The plane of
acquisition is perpendicular to the antenna axis. (b) The contribution
of the effective velocity contrast $A^{*}_{c}$; (c) the contribution
of the effective impedance contrast $A^{*}_{Z}$. Zero-offset data
record only the effective impedance contribution.}

\label{fig: simu sable}
\end{figure}

Introducing the expressions for the incident electric field~(\ref{eq:
  ricker}), and for the anisotropic antenna radiation pattern in
equation~(\ref{eq: champ dif}), a synthetic radargram is
computed~(Figure~\ref{fig: simu sable}a), assuming that the antennas
are perpendicular to the plane of acquisition. Note that other planes
could have been chosen as the field is diffracted in 3 dimensions. The
geometric dispersion is included in equations~(\ref{eq: ricker})
and~(\ref{eq: champ dif}). Then the contributions of the impedance
contrast and the velocity contrast are calculated (Figures~\ref{fig:
  simu sable}b and ~\ref{fig: simu sable}c).

\begin{figure}
\centerline{\includegraphics*[width=8cm]{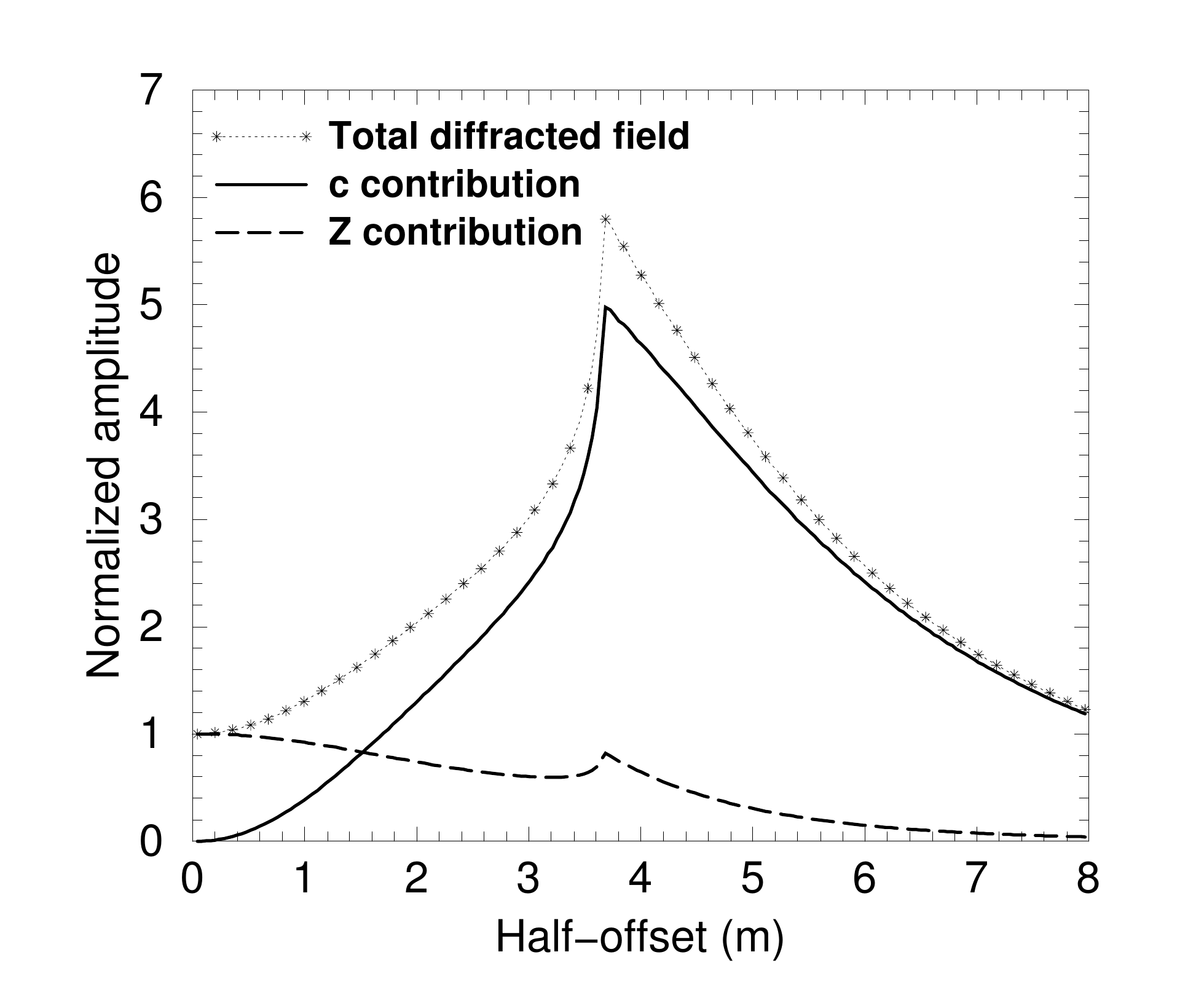} \hfill
  \includegraphics*[width=8cm]{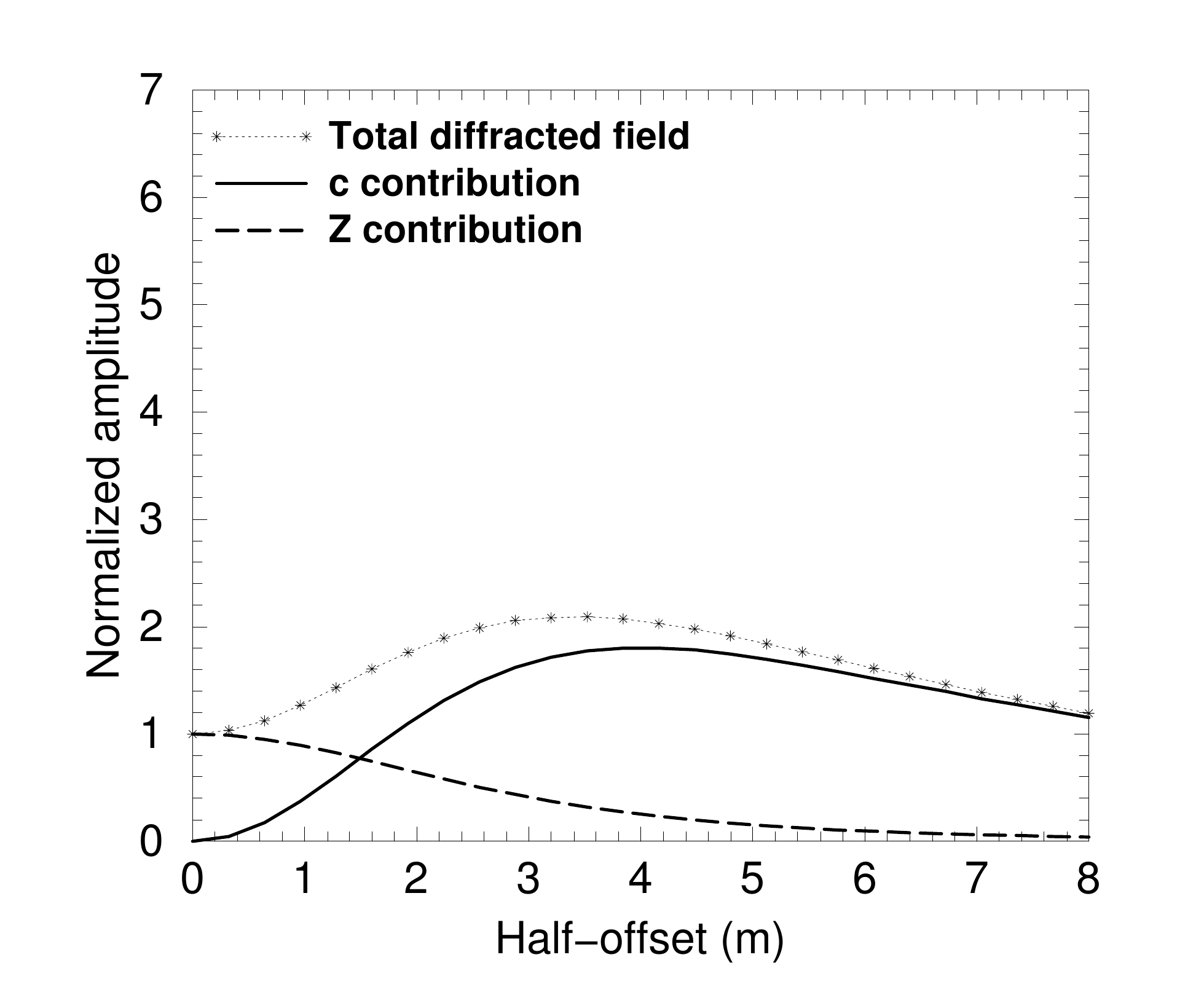}}
\hspace{2.5cm} a) \hspace{7cm} b)
\vspace{5mm}

\caption{(a) Trace to trace maximum amplitude, normalized to the
  maximum amplitude of the zero-offset trace, of the radargrams in
  Figure~\ref{fig: simu sable} recomputed with 4 times as many traces.
  For clarification, amplitudes have been recalculated and displayed
  in (b) for omnidirectional source and receiver.  In this numerical
  example, $A^*_Z$ is 7 times less than $A^*_c$, so, even with the
  geometric dispersion, the amplitude first increases with the offset.
  Large offset data contain mainly information on velocity contrasts.}

\label{fig: simu amp}
\end{figure}

Trace to trace maximum amplitudes, normalized to the maximum amplitude
of the zero-offset trace, are presented in Figure~\ref{fig: simu amp}.
Figure~\ref{fig: simu amp}a results from the calculation using antenna
radiation pattern expressions from Engheta \shortcite{engheta82}. In
Figure~\ref{fig: simu amp}b, we use omnidirectional antenna radiation
patterns to show that the amplitude peaks at 3.8 m are caused by the
radiation pattern of the point dipole antennas. However, in our
example, $A_Z^*$ ($\approx -1.5\,10^{-5}\ \m^3$) is 7 times less than
$A_c^*$ ($\approx -11.1\,10^{-5}\ \m^3$), so, even with the
geometrical dispersion, the amplitude is, at first, increasing when
the offset between the two antennas increases. To conclude this
numerical application, Figure~\ref{fig: simu amp} emphasizes that,
independent from antenna radiation patterns, the electric field
recorded at zero offset depends only on the effective impedance
contrast, and, the effective velocity contrast contribution dominates
for half-offsets larger than 2 m in this case.

\section{Conclusions}

We have studied a point diffractor in a homogeneous, linear,
isotropic, low-conductivity medium, using different parameterizations
and have made several observations useful for \GPR acquisition. First,
by studying the amplitude of the field diffracted by a small anomaly,
it has been shown that magnetic variations of the ground cannot be
ignored without overwhelming justification.

In the far field approximation, a point logarithmic effective
dielectric permittivity anomaly acts as a small electric dipole. A
point logarithmic effective magnetic permeability anomaly acts as a
small magnetic dipole. These behave differently. Therefore, dielectric
and magnetic effects can be theoretically distinguished with \GPR
surface data, by varying the antenna orientations. However, it remains
difficult to apply in real life because of the lack of precision in
antenna directionality.

Instead of parameterizing a medium with dielectric and magnetic
parameters, we may use effective electromagnetic impedance and
effective electromagnetic velocity. Radiation patterns show that this
parameterization is very useful for the inverse problem. Indeed, the
effective impedance controls the amplitude reflected at small offsets,
whereas the effective velocity controls the amplitude reflected at
large offsets. It implies that a multi-offset data acquisition allows
a multi-parameter measurement. A radar image obtained in monostatic
mode is a direct image of the effective impedance contrasts. A radar
image obtained with a large offset between the source and the receiver
contains information about effective velocity contrasts.

In terms of 3D modeling, the equations developed in this paper provide
a valid starting point given the limitations outlined previously.
However a good model should take into account the near-field, and a
conductive and dispersive medium. This can be done considering the
near-field term in the expression of the Green's tensor and staying in
the frequency domain. This is the subject of future work.

\section{Acknowledgments}

This work was supported by the Bureau de Recherche G\'eologique et
Mini\`ere (Orl\'eans). Comments and suggestions by John Scales, Gary
Olhoeft, and the reviewers are gratefully acknowledged.


\bibliography{mabib}

\end{document}